\documentclass[a4paper,nofootinbib,twocolumn]{revtex4-1}

\usepackage{epsfig}
\usepackage{amsmath,amsfonts,amssymb}
\usepackage{t1enc}

\providecommand{\openone}{\leavevmode\hbox{\small1\kern-3.8pt\normalsize1}}

\newcommand{\Vl}{V_L}
\newcommand{\Vr}{V_R}
\newcommand{\gl}{g_L}
\newcommand{\gr}{g_R}

\newcommand{\fp}{F_+}

\newcommand{\fz}{F_0}
\newcommand{\Apm}{A_\pm}
\newcommand{\Ap}{A_+}
\newcommand{\Am}{A_-}

\begin{document}

\title{Constraints on the $Wtb$ vertex from early LHC data}

\author{J. A. Aguilar--Saavedra}
\affiliation{Departamento de Física Teórica y del Cosmos and CAFPE, \\
Universidad de Granada, E-18071 Granada, Spain}
\author{N. F. Castro, A. Onofre}
\affiliation{LIP, Departamento de Fisica, Universidade do Minho, P-4710-057 Braga, Portugal}

\begin{abstract}
We use the recent measurements of top quark decay asymmetries in ATLAS and the $t$-channel single top cross section in CMS to set the first combined LHC limits on the $Wtb$ vertex. This combination allows to obtain much better limits than the separate measurements. The resulting constraints are comparable, although still weaker, than the ones obtained using Tevatron data with much more statistics.
\end{abstract}


\maketitle

\section{Introduction}

The arrival of the first LHC data starts a new era in precision studies of the top quark properties. Even with the few statistics collected in 2010, the CMS and ATLAS collaborations have been able to present useful, and often competitive, measurements and limits on new physics related to the top quark. In particular, a first glance at the top decay has been given through the study of $W$ helicity fractions and related observables~\cite{Wh-ATLAS}. Production cross sections have been measured both for $t \bar t$ pairs~\cite{Khachatryan:2010ez,Aad:2010ey} and for single top quarks~\cite{st-CMS,st-ATLAS}.
The aim of this note is to provide a first combined limit on new physics contributions to the $Wtb$ vertex using top decay observables and single top cross sections measured at LHC, to show the important improvement brought by this combination with real LHC data.

We work in an effective field theory framework~\cite{Buchmuller:1985jz} to parameterise the effect of heavy new physics on the $Wtb$ interaction. Upon electroweak symmetry breaking, the most general $Wtb$ vertex including corrections from dimension-six gauge invariant operators is~\cite{AguilarSaavedra:2008zc}
\begin{eqnarray}
\mathcal{L}_{Wtb} & = & - \frac{g}{\sqrt 2} \bar b \, \gamma^{\mu} \left( \Vl
P_L + \Vr P_R
\right) t\; W_\mu^- \nonumber \\
& & - \frac{g}{\sqrt 2} \bar b \, \frac{i \sigma^{\mu \nu} q_\nu}{M_W}
\left( \gl P_L + \gr P_R \right) t\; W_\mu^- + \mathrm{H.c.} \,, \notag \\
\label{ec:lagr}
\end{eqnarray}
with $\Vl$, $\Vr$, $\gl$, $\gr$ complex dimensionless constants.
In the Standard Model (SM), the $Wtb$ vertex in Eq.~(\ref{ec:lagr}) reduces to $\Vl = V_{tb} \simeq 1$ at the tree level. Corrections to this coupling, as well as non-zero
anomalous couplings $\Vr$, $\gl$ and $\gr$ can be generated by heavy new physics.  These anomalous couplings can be probed in top decays through the measurement of the $W$ helicity fractions~\cite{Kane:1991bg} or directly related observables~\cite{delAguila:2002nf,AguilarSaavedra:2006fy}. In the production, they can be probed with a measurement of the single top cross sections~\cite{Boos:1999dd,Najafabadi:2008pb,AguilarSaavedra:2008gt}. The combination of both production and decay observables to constrain the $Wtb$ vertex has already been discussed extensively in the literature~\cite{Chen:2005vr,AguilarSaavedra:2008gt,AguilarSaavedra:2010nx,Zhang:2010dr}. In this note we follow Refs.~\cite{AguilarSaavedra:2008gt,AguilarSaavedra:2010nx} and use the dedicated code {\sc TopFit} to perform a combination of top decay observables and single top cross sections, measured either at LHC or at Tevatron, to obtain constraints on anomalous $Wtb$ couplings.

\section{Collider observables for $Wtb$ anomalous couplings}
\subsection{Top decay}

At Tevatron, the $W$ helicity fractions in the decay $t \to Wb$ have been precisely measured by  both the CDF and D0 collaborations. We use the latest results for the semileptonic $t \bar t$ decay channel from CDF~\cite{Aaltonen:2010ha}
\begin{align}
& \fz = 0.88 \pm 0.11\;\text{(stat)} \pm 0.06\;\text{(syst)} \,, \notag \\
& \fp = -0.15 \pm 0.07\;\text{(stat)} \pm 0.06\;\text{(syst)} \,, 
\end{align}
with a correlation coefficient $\rho = -0.59$, assuming a top quark mass $m_t = 175$ GeV. The combination of semileptonic and dilepton decay channels from D0 gives~\cite{Abazov:2010jn},
\begin{align}
& \fz = 0.669 \pm 0.078\;\text{(stat)} \pm 0.065\;\text{(syst)} \,, \notag \\
& \fp = 0.023 \pm 0.041\;\text{(stat)} \pm 0.034\;\text{(syst)} \,, 
\end{align}
with $\rho = -0.83$, assuming a top quark mass $m_t = 172.5$ GeV. We do not include a CDF measurement in the dilepton decay channel~\cite{CDFdilep} (with a smaller sensitivity) because correlations with the limit from the semileptonic channel are not known to us. We also assume that correlations among the systematic uncertainties present in both experiments can be neglected.

For LHC, rather than using the helicity fraction themselves, we use angular asymmetries $\Apm$ on the $\cos \theta_\ell^*$ distribution, where $\theta_\ell^*$ is the angle between the charged lepton momentum in the $W$ rest frame and the $W$ momentum in the top quark rest frame. These asymmetries are~\cite{AguilarSaavedra:2006fy},
\begin{equation}
\Apm = \frac{N(\cos \theta_\ell^* > z_\pm) - N(\cos \theta_\ell^* < z_\pm)}{N(\cos \theta_\ell^* > z_\pm) + N(\cos \theta_\ell^* < z_\pm)} \,,
\end{equation}
where $N$ stands for the number of events and $z_\pm = \mp (2^{2/3}-1)$. They are better suited than the helicity fractions for setting constraints on anomalous $Wtb$ couplings with low statistics, and have been measured by ATLAS using approximately 35 pb$^{-1}$ of data, in $t \bar t$ production with decays into the semileptonic channel,
\begin{align}
& \Ap = 0.50 \pm 0.10\;\text{(stat)} \pm 0.06\;\text{(syst)} && (e) \,, \notag \\
& \Am = -0.85 \pm 0.07\;\text{(stat)} \pm 0.05\;\text{(syst)} && (e) \,, \notag \\
& \Ap = 0.50 \pm 0.08\;\text{(stat)} \pm 0.04\;\text{(syst)} && (\mu) \,, \notag \\
& \Am = -0.87 \pm 0.04\;\text{(stat)} \pm 0.03\;\text{(syst)} && (\mu) \,.
\end{align}
The correlation between $\Ap$ and $\Am$ is $\rho = 0.16$ for each decay channel $e,\mu$.
The top quark mass is taken as $m_t = 172.5$ GeV.

\subsection{Single top production}

The CDF and D0 collaborations have provided evidence for single top production at Tevatron.
In our fits we use the combined measurement from both experiments~\cite{Group:2009qk}
of the cross sections for $s+t$ channel production,
\begin{equation}
\sigma_{s+t} = 2.76^{+0.58}_{-0.47}~\text{pb} \,,
\end{equation}
which assumes $m_t = 170$ GeV. Separate measurements for $s$- and $t$-channel production are available but their precision is lower and the resulting constraints on anomalous $Wtb$ couplings are weaker. We ignore possible (likely small) correlations between the systematic uncertainties for this cross section measurement and helicity fractions.

At LHC, both CMS and ATLAS have provided measurements of the $t$-channel single top cross section at 7 TeV~\cite{st-CMS,st-ATLAS},
\begin{align}
& \sigma_t = 83.6 \pm 30.0~\text{pb} && \text{(CMS)} \,, \notag \\
& \sigma_t = 53^{+46}_{-36}~\text{pb} && \text{(ATLAS)} \,,
\end{align}
taking $m_t = 172.5$ GeV.
The former measurement has a better precision while the latter has the central value closest to the SM next-to-leading order cross section $\sigma_t = 61.9 \pm 2.7~\text{pb}$~\cite{Schwienhorst:2010je}.\footnote{This value assumes $m_t = 173$ GeV but the difference with $m_t = 172.5$ GeV, used in the fits, is negligible compared to the experimental uncertainty.}
Hence, both measurements provide very similar limits on $Wtb$ anomalous couplings, and we use the one from CMS to avoid possible systematic uncertainty correlations with the top decay asymmetries (measured by ATLAS) which would have to be addressed in detail otherwise.

\section{Limits}

Limits on the $Wtb$ couplings in Eq.~(\ref{ec:lagr}) are set by using {\sc TopFit} which implements the analytical expressions of $W$ helicity fractions and related observables~\cite{AguilarSaavedra:2006fy}, as well as the single top cross sections~\cite{AguilarSaavedra:2008gt}, in terms of $\Vl$, $\Vr$, $\gl$ and $\gr$. For a top quark mass $m_t = 172.5$ GeV (as assumed for the recent LHC measurements), $M_W = 80.4$ GeV and $m_b = 4.8$ GeV, the SM tree-level prediction for helicity fractions is $\fz = 0.696$, $\fp = 3.8 \times 10^{-4}$, and for the related asymmetries $\Ap = 0.543$, $\Am = -0.841$. We have explicitly checked that the variation in the limits when using $m_t=170,175$ GeV but keeping the same experimental values is minimal.
QCD corrections~\cite{Czarnecki:2010gb} are much smaller than the experimental uncertainty, and are therefore ignored.
The SM single top cross sections for Tevatron and LHC are taken as $\sigma_{s+t} = 2.86 \pm 0.36$ pb~\cite{Sullivan:2004ie}, $\sigma_t = 61.9 \pm 2.7~\text{pb}$~\cite{Schwienhorst:2010je}, respectively. Corrections to these cross section from anomalous couplings are evaluated using {\sc Protos}~\cite{AguilarSaavedra:2008gt}.

Due to the limited single top statistics ---even at Tevatron--- and the need for further independent observables still to be measured, a global fit to the general complex $Wtb$ vertex, as proposed in Ref.~\cite{AguilarSaavedra:2010nx}, is not possible. Instead, we focus here on subsets of couplings, assuming for the rest their SM value. This approach is, albeit not the most general, perfectly consistent because the different couplings arise from different gauge-invariant operators. We also assume that anomalous couplings are real ($\Vl$ can be taken real and positive by definition). It must be pointed out, in addition, that we ignore possible four-fermion contributions to the $t$-channel single top cross sections~\cite{Cao:2007ea,AguilarSaavedra:2010zi}. In an effective operator framework, there are several dimension-six four-fermion operators which potentially contribute to this process. Being gauge-invariant, their contribution can be ignored without losing internal consistency; we remark again that this approach is not the most general one but it is necessary with the presently available data.

\begin{figure*}[t]
\begin{center}
\begin{tabular}{ccc}
\epsfig{file=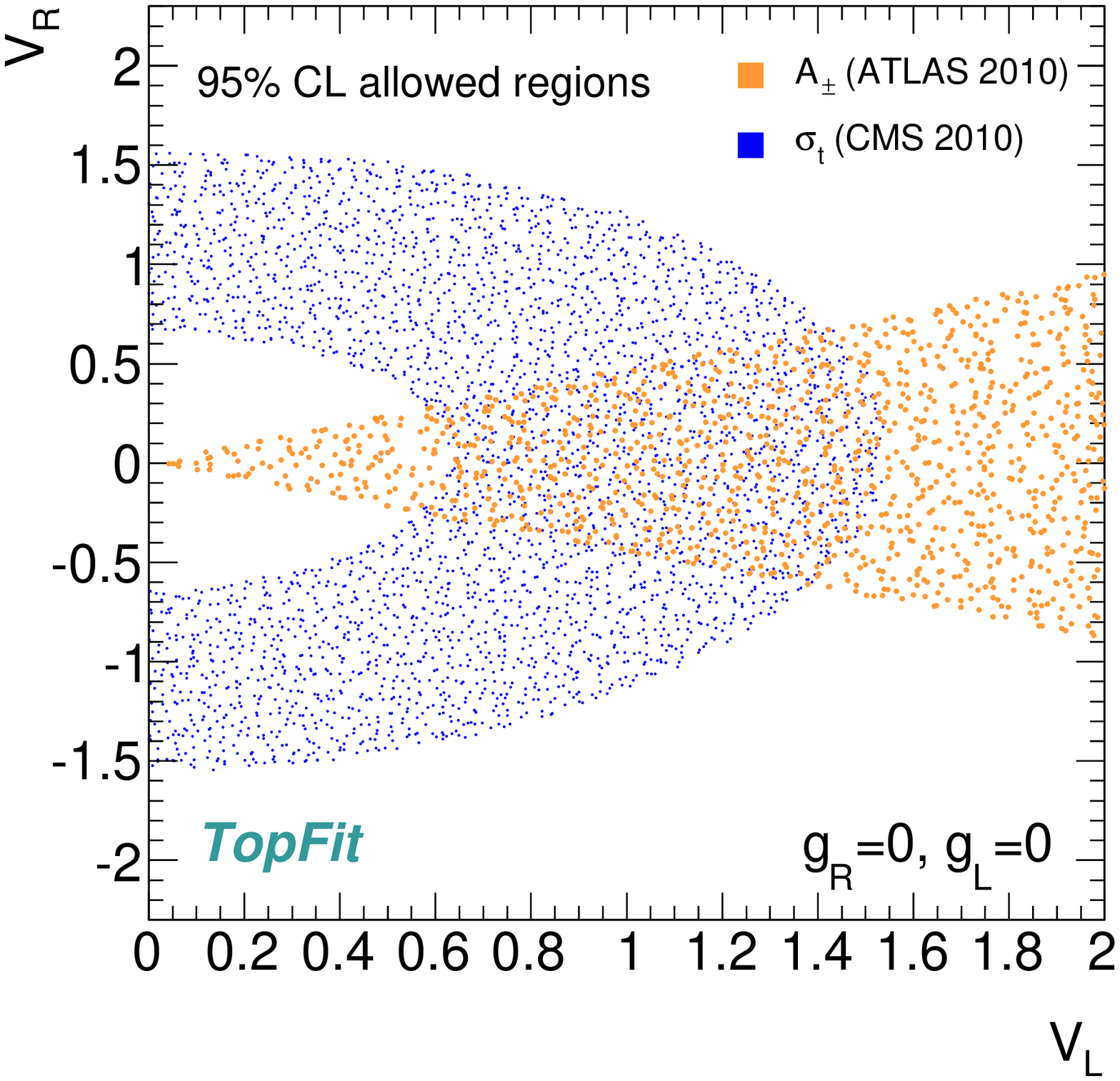,height=6cm,clip=} & \quad &
\epsfig{file=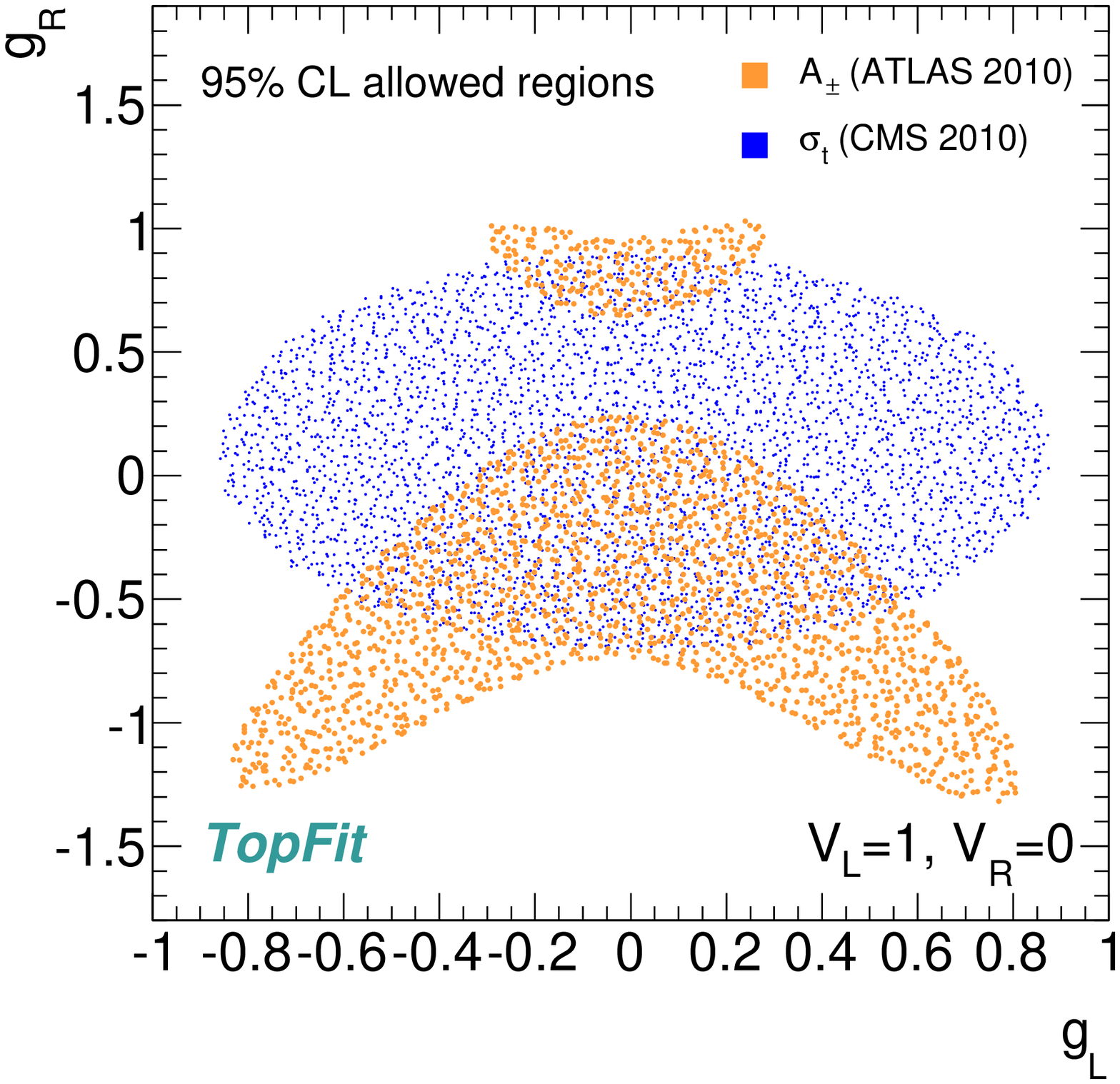,height=6cm,clip=}
\end{tabular}
\caption{LHC limits on $\Vl,\Vr$ (left) and $\gl,\gr$ (right) from top decays and single top production.}
\label{fig:lim1}
\end{center}
\end{figure*}

We present in Fig.~\ref{fig:lim1} representative limits from single top production measured at CMS and top decay asymmetries measured in ATLAS. The left panel corresponds to the limits on $(\Vl,\Vr)$ assuming $\gl = \gr = 0$. The complementarity of both measurements is beautifully depicted here: the intersection of the arc-shaped region from $\sigma_t$ and the triangle from $\Apm$ gives a much more stringent limit than the separate measurements.
\begin{figure*}[t]
\begin{center}
\begin{tabular}{ccc}
\epsfig{file=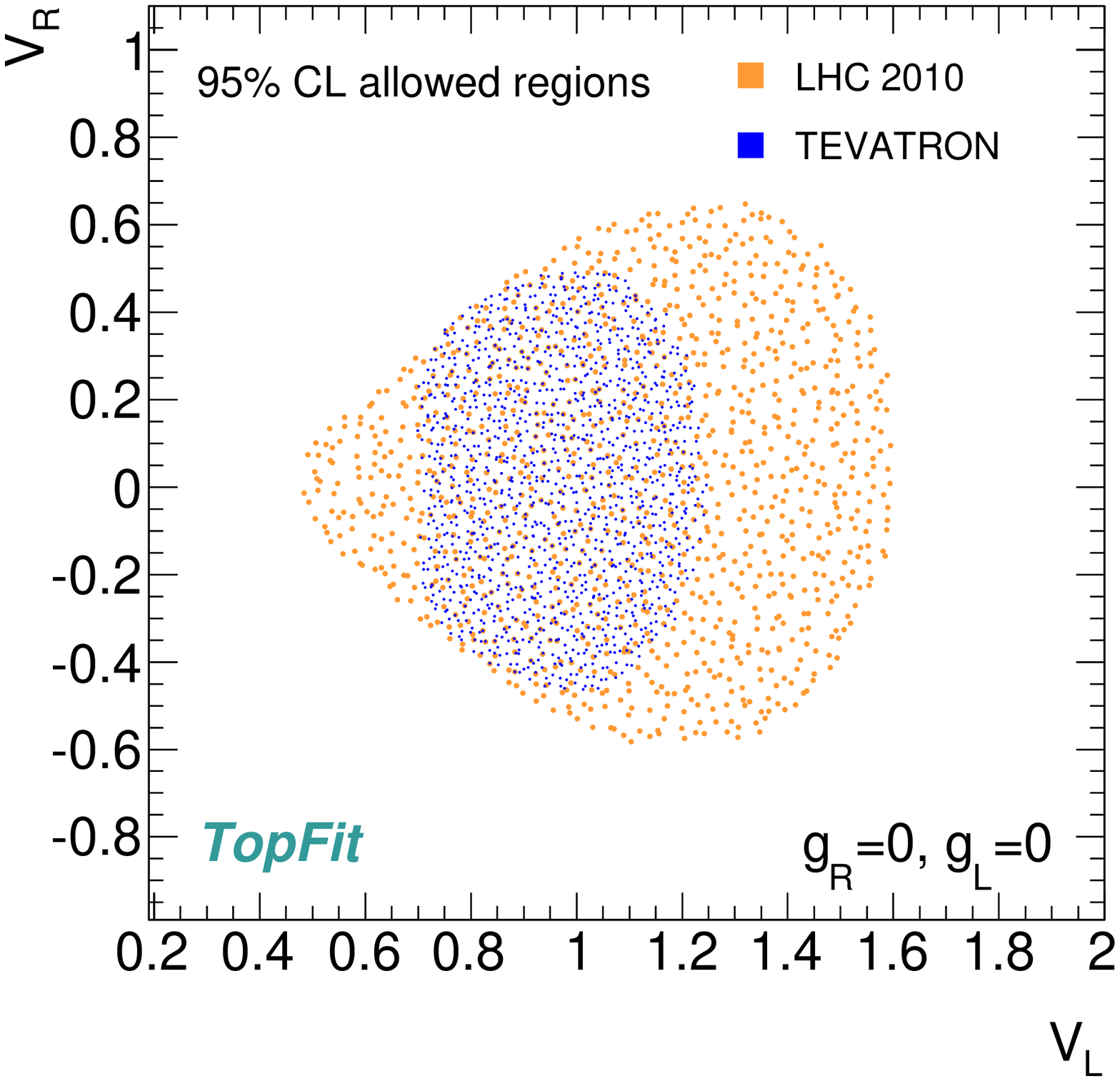,height=6cm,clip=} & \quad &
\epsfig{file=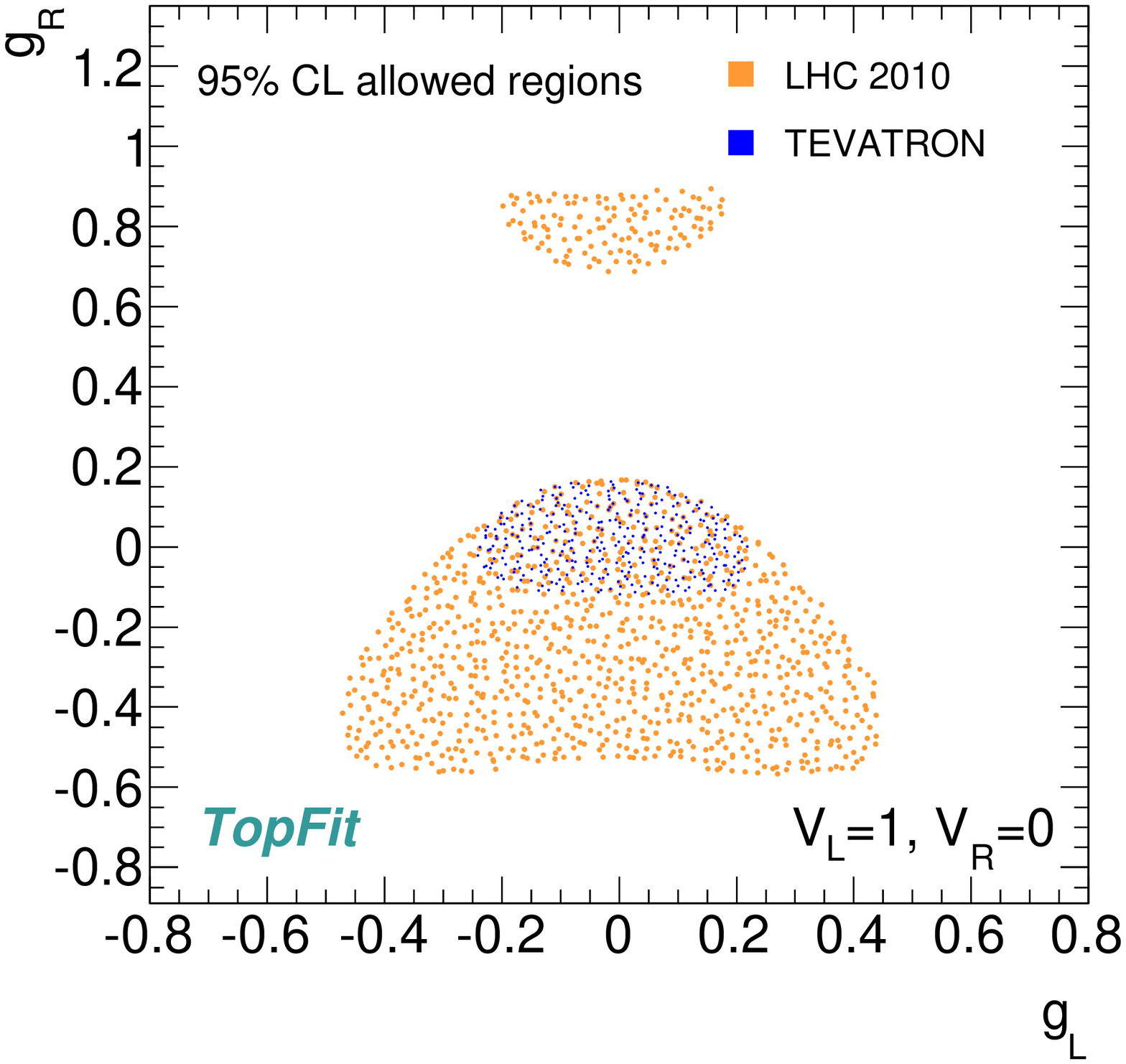,height=6cm,clip=}
\end{tabular}
\caption{Comparison of LHC and Tevatron combined limits on $\Vl,\Vr$ (left) and $\gl,\gr$ (right).}
\label{fig:lim2}
\end{center}
\end{figure*}
On the right panel we present the limits on $(\gl,\gr)$ assuming $\Vl = 1$, $\Vr = 0$. Again, the combination of both measurements is very powerful and almost removes the large $\gr$ region present in the limit from $\Apm$. The resulting combined LHC limits are shown in Fig.~\ref{fig:lim2}, including also the analogous ones from Tevatron. We observe that these early LHC limits are not too far from the Tevatron ones, despite the still small statistics available. And they will readily improve in the near future with the new data being collected by the LHC experiments with a quickly growing integrated luminosity.

\section{Conclusions}

In this note we have used the measurements of top decay asymmetries and single top production cross sections from ATLAS and CMS, respectively, in order to obtain the first combined limits on the $Wtb$ vertex using LHC data. We have shown, with few selected examples, the great benefit of such combination already at the early LHC phase, when top decay observables are still dominated by statistics and the relative error of the single top cross section is above 30\%. We advocate for the implementation of these combined limits, not only within a single experiment but including all available data from CMS and ATLAS, to provide constraints as stringent as possible on anomalous $Wtb$ couplings.
\vspace{0.5cm}

\section*{Acknowledgements}

This work has been partially supported by FCT (project CERN/FP/116397/2010 and grant SFRH/BPD/63495/2009), CRUP (Ac\c{c}\~ao integrada Ref. E 2/09),
MICINN (FPA2010-17915 and HP2008-0039) and Junta de Andaluc\'{\i}a (FQM 101 and FQM 437).

\end{document}